\pdfoutput=1
\documentclass[]{spie}  

\newcommand{\bra}[1]{{\langle #1 \vert}} 		
\newcommand{\ket}[1]{{\vert #1 \rangle}} 		

\usepackage{amsmath,amsfonts,amssymb}
\usepackage{graphicx}
\usepackage[colorlinks=true, allcolors=blue]{hyperref}

\title{Theoretical comparison of quantum and classical illumination for simple detection-based LIDAR}

\author{Richard J. Murchie, Jonathan D. Pritchard, and John Jeffers}

\affil{Department of Physics, SUPA, University of Strathclyde, John Anderson Building, 107 Rottenrow East, Glasgow G4 0NG, United Kingdom}

\begin{document} 
\maketitle

\begin{abstract}
Use of non-classical light in a quantum illumination scheme provides an advantage over classical illumination when used for LIDAR with a simple and realistic detection scheme based on Geiger-mode single photon detectors. Here we provide an analysis that accounts for the additional information gained when detectors do not fire that is typically neglected and show an improvement in performance of quantum illumination. Moreover, we provide a theoretical framework quantifying performance of both quantum and classical illumination for simple target detection, showing parameters for which a quantum advantage exists. Knowledge of the regimes that demonstrate a quantum advantage will inform where possible practical quantum LIDAR utilising non-classical light could be realised. 
\end{abstract}


\section{INTRODUCTION}
\label{sec:intro}  
Simple detection-based LIDAR consists of sending a signal state of light out into an environment that may, or may not, contain a target object, then detecting possible reflected light. Any light reflected to the detector will provide information about the presence of a possible target object. However, when the state of light has a low mean photon number and there is high environmental background noise, accurate inference of the presence of an object is challenging. This problem amounts to discriminating between two states, one containing the reflected signal and noise and the other only noise, so it can be expressed in terms of quantum state discrimination. We determine if an object is present or not by attempting to discriminate the possible states incident on our detector system. The more distinguishable these states are, the more quickly the presence of an object can be recognised or excluded. Quantum illumination exploits non-classically correlated optical modes as the light source to perform object detection\cite{PhysRevLett.101.253601,S.Barzanjeh}, offering a fundamental advantage over classical light sources due to the enhanced distinguishability of non-classical quantum states. Quantum illumination can thus yield improved target discrimination even in noisy quantum channels\cite{Sacchi,Sacchi_2005}, however the exact measurement protocol required to achieve the maximal enhancement in sensitivity is currently unknown\cite{Lloyd2008}. This scheme can be implemented in various ways. If one mode (conventionally ``the idler'') is stored locally until a return signal appears at the detector the two modes may be detected in combination to obtain the detection advantage. This is challenging if it requires interference and hence a phase lock between the idler and signal beams, so it is impractical outside a laboratory at optical frequencies. A more practical method entails locally measuring the idler, then using this measurement to condition the signal beam, which is sent to interrogate the target. The hope is that the conditioned signal beam will have enhanced detection probability. Quantum illumination has been shown to have advantages over classical illumination for object detection both experimentally\cite{England2019} and theoretically\cite{HYang}, when simple detection is used for both the signal and the idler.

In this paper we extend the analysis of quantum illumination for 
object detection with simple detectors, with a focus of using all of the detector information available. Moreover, we develop a distinguishability measure to assess the performance of illumination in a simple LIDAR system, in order to demonstrate where a quantum advantage exists.
The paper is organised as follows. In Section \ref{simlidmod} we develop a model for both a classical and a quantum illumination-based LIDAR system with Geiger-mode avalanche photodiode detectors. We provide simple models for the classical and quantum sources, for the target object and for the detection of the signal. In Section \ref{odcd} we describe the theory of object detection with click detectors and compare the click count distributions for both classical and quantum illumination of a target object. In Section \ref{llt} we describe the application of the log-likelihood test to the distributions of click counts and use this to show that quantum illumination produces a much greater difference in log-likelihood values than classical when it is used to illuminate a possible target. In Section \ref{dm} we introduce a distinguishability measure based on the integrated log-likelihood statistics for determining the improvements expected by using a heralded light source rather than a classical one. In the final section we discuss our findings. 

\section{SIMPLE LIDAR MODEL}
\label{simlidmod}
The basic problem consists of the recognition of a possible target object by sending states of light in its direction and detecting any light that reflects off. We compare how easy it is to either detect, or rule out the presence of, the target if we send either classical or quantum states. In order to simulate the above problem, we construct a model for both a quantum illumination and a classical illumination-based LIDAR system\cite{Rohde2006}. Both the classical and quantum source will have the same signal strength, characterised by the mean photon number of the light source. Additionally any model parameter that appears in both quantum and classical illumination systems is equal, which will facilitate later comparison.

As we wish to model experimentally feasible detectors, signal analysis for object detection will arise from use of measurement operators (POVMs) that model Gieger-mode photodetectors. In order to account for the information gained using a Gieger-mode photodetector in a single shot, we model measurements of the state incident on the signal detector for classical illumination and for quantum illumination, which allows us to calculate the probability of no-click in each case. The expectation value of a quantum state and a no-click POVM yields the probability of a no-click event\cite{Barnett}.
\begin{figure}
    \centering
    \includegraphics[width=\linewidth]{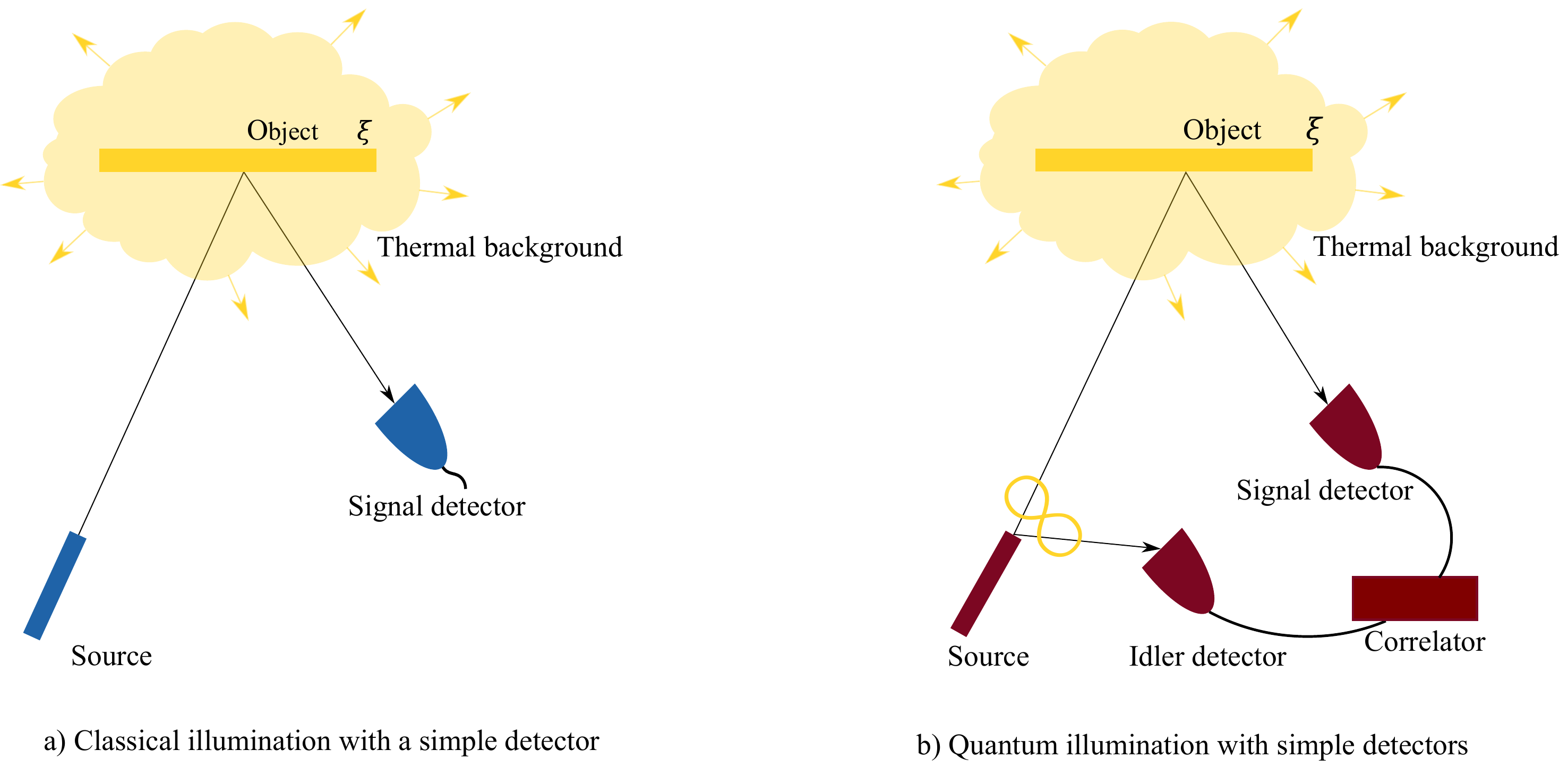}
    \caption{Schematic showing LIDAR using classical illumination with a simple detector (a) and quantum illumination with simple detectors (b), where the signal and idler beam is photon number correlated. The presence of an object is defined by object reflectivity $\xi$.
    }
    \label{fig:my_label}
\end{figure}
\subsection{Classical light source} 
We use as a classical light source a single-mode thermal state with mean photon number $\bar{n}_{\textrm{S}}$ given by\cite{Mandel1995},

\begin{equation}
    \hat{\rho}_{\textrm{th}}=\sum_{n=0}^\infty\frac{\bar{n}_{\textrm{S}}^n}{(1+\bar{n}_{\textrm{S}})^{n+1}}\ket{n}\bra{n}, 
\end{equation}
where $\ket{n}$ is the photon number state. This is not the optimal classical state to send to the target. A state with a Poissonian photon number distribution, such as what is produced by a laser above threshold, would provide slightly better discrimination, but the differences between the statistics obtained using a single-mode coherent state and a single-mode state with thermal statistics are small in the regime considered here\cite{Scully}. Our classical light source has a Bose-Einstein photon number distribution, which is the same as the background noise of the environment, permitting the classical light source to be useful in situations that require covertness.

\subsection{Quantum light source} 
Our quantum light source is a pair of beams such as the signal and idler beams produced by parametric down-conversion or four-wave mixing, where a pump laser mode is transformed into a pair of modes with pulse trains correlated in photon number \cite{Bachor2004}. We assume that the idler is measured and the signal is sent to interrogate the object. The source produces a pulsed two-mode squeezed vacuum (TMSV), a quantum state characterised by a mean photon number $\bar{n}_{\textrm{S}}$, with strong correlations in photon number between the two modes (or beams),
\begin{equation}
    \hat{\rho}_{\textrm{TMSV}}=\sum_{n=0}^\infty\frac{\bar{n}_{\textrm{S}}^n}{(1+\bar{n}_{\textrm{S}})^{n+1}}\ket{n}_{\textrm{S}}\ket{n}_{\textrm{I}}\bra{n}_{\textrm{I}}\bra{n}_{\textrm{S}},
\end{equation}
where $\ket{n}$ is the photon number state, $\textrm{S}$ denotes the signal mode and $\textrm{I}$ denotes the idler mode. In low mean photon number regimes, any light produced is predominately a correlated pair of single photons.

Any measurement on the idler mode conditions the signal state sent to the target. Furthermore, the quantum source, ignoring the idler, is identical to a single-mode thermal source. It is therefore more difficult to distinguish from the background than a single-mode coherent state of the same mean photon number. In order to exploit the non-classical correlations, the idler detector results must be used. We model this measurement based on a Geiger-mode photodetector which can only provide either a click-event or a no-click event in each shot of the experiment. If the idler detector does not produce dark counts the no-click POVM is 
\begin{equation}
    \hat{\pi}_0^{\textrm{I}}=\sum_{n=0}^\infty\left(1-\eta_{\textrm{I}}\right)^n\ket{n}\bra{n},
\end{equation} 
where $\ket{n}$ is the photon number state and $\eta_{\textrm{I}}$ is the quantum efficiency of the idler detector. Dark counts can be included via a thermal mean photon number $\bar{n}_{\textrm{B,I}}$ or simply a firing probability when no light is incident on the detector. The other measurement result is determined by the click POVM for the idler detector, $\hat{\pi}_1^{\textrm{I}}=\hat{1}-\hat{\pi}_0^{\textrm{I}}$ and this complementary nature between click and no-click POVM holds in general. Thus the probability of an idler-firing event is $P(1)_{\textrm{I}}=\text{Tr}(\hat{\rho}_{\textrm{TMSV}}\hat{\pi}_1^{\textrm{I}})$. The signal detector works in the same way as the idler detector.

The quantum states incident on the signal detector after measuring an idler firing event or an idler no-firing event are respectively
\begin{align}
    \hat{\rho}_{\textrm{S,1}}&=\frac{\text{Tr}_{\textrm{I}}(\hat{\rho}_{\textrm{TMSV}}\hat{\pi}_1^{\textrm{I}})}{\text{Tr}(\hat{\rho}_{\textrm{TMSV}}\hat{\pi}_1^{\textrm{I}})}, \\ \hat{\rho}_{\textrm{S,0}}&=\frac{\text{Tr}_{\textrm{I}}(\hat{\rho}_{\textrm{TMSV}}\hat{\pi}_0^{\textrm{I}})}{\text{Tr}(\hat{\rho}_{\textrm{TMSV}}\hat{\pi}_0^{\textrm{I}})},
\end{align}
where $\text{Tr}_{\textrm{I}}$ is the partial trace over the idler mode. This conditioning of the state that is incident on the target, due to our local measurements on the idler detector is what gives quantum illumination its distinct advantage over classical illumination.
\subsection{Target object model}
There will be a difference in the intensity of incident light upon the signal detector depending on whether an object is present or not as, if an object is present, there will be light reflected onto the signal detector from our light source. Object detection is a quantum state discrimination problem — distinguishing between the incident quantum state when an object is present or not. Environmental background noise, detector inefficiencies and sub-optimal object reflectivity all influence the system, and will cause discrimination between the possible quantum states to become more difficult. In both classical and quantum schemes in order to model the absence of an object and hence the signal being lost to the environment we set the object reflectivity parameter to be $\xi=0$.
\section{OBJECT DETECTION WITH CLICK DETECTORS}
\label{odcd}
As determining the presence of an object is a quantum state discrimination problem it is worthwhile to mention that an optimal measurement exists. While the exact optimal measurement is unknown because its POVM is, in general, difficult to calculate and realise, calculable bounds exist\cite{Helstrom1969}. The bounds demonstrate that the quantum illumination states are more distinguishable than the corresponding states from a classical illumination-based LIDAR system\cite{Audenaert2007}. Following on from the discussion on POVMs in Section \ref{simlidmod}, we extend this to include other aspects of our model. As our target object and our inefficient detector models each include an attenuation parameter (reflectivity or quantum efficiency) and a thermal background (the target environment and the detector dark noise) we use a combined POVM for the reflection and detection that accounts for these aspects of our LIDAR model: sub-optimal object reflectivity, environmental background noise, and detector inefficiencies. Then probabilities of click and no-click events on the detectors for both classical and quantum illumination is all of the information we can calculate in this simple detection LIDAR scheme.

\subsection{Classical illumination} 
The signal no-click POVM element is $\hat{\pi}_0$, which includes the background noise and the possible target object. This is similar to the no-click POVM for the idler detector, with the adjustment that the product of object reflectivity $\xi$ and quantum efficiency of the signal detector $\eta_{\textrm{S}}$ is the attenuating factor. Moreover, the POVM contains the mean photon number of the environmental background noise $\bar{n}_{\textrm{B,S}}$ instead of idler detector dark noise mean photon number $\bar{n}_{\textrm{B,I}}$. For each shot the no-click probability is
\begin{equation}
\text{P(0)}_{\textrm{CI}}=\text{Tr}(\hat{\rho}\hat{\pi}_0),
\end{equation}where $\hat{\rho}$ is the classical light source, and if the object is absent the object reflectivity $\xi=0$.

\subsection{Quantum illumination} 

The information available for a quantum illumination-based LIDAR system differs from a classical illumination-based LIDAR system as there are two detectors: the idler detector and the signal detector. Following from this, we seek to gain information from all possible events, not just when the idler detector fires. The existing literature on quantum illumination-based object detection relies only upon idler firing events and overlooks idler not firing. In order to account for all possible events we seek the probability of no-click on the signal detector when the idler fires {\it and} when it does not. When an object is present, the probability of no-click on the signal detector when the idler detector fires and does not fire are, respectively
\begin{equation}
\text{P(0)}_{\textrm{S,1}}=\text{Tr}(\hat{\rho}_{\textrm{S,1}}\hat{\pi}_0) \; \; \text{and} \; \; \text{P(0)}_{\textrm{S,0}}=\text{Tr}(\hat{\rho}_{\textrm{S,0}}\hat{\pi}_0),    
\end{equation} where $\hat{\rho}_{\textrm{S,1}}$ is the conditioned state at the signal detector after an idler-firing event, $\hat{\rho}_{\textrm{S,0}}$ is the conditioned state at the signal detector after an idler not firing event and $\hat{\pi}_0$ is the signal no-click POVM element. Whereas, when no object is present, the probability of no-click on the signal detector when the idler detector fires and does not fire,
\begin{equation}\text{P(0)}_{\textrm{QI,NO}}=\text{Tr}(\hat{\rho}\hat{\pi}_0),\end{equation} where $\hat{\rho}$ is either conditioned signal state and the object reflectivity $\xi=0$. While the main advantage of quantum illumination is due to the increase in conditioned photon number when the idler detector fires, there is advantage to be gained from considering the states sent when the idler does not fire, as we shall see later.
\subsection{Click count distributions}
There is a limit to the amount of information that we can gain from a single shot of the experiment. In order to remedy this problem we perform quantum hypothesis testing where many shots of the experiment generate a probability distribution of cumulative click counts recorded by the signal detector\cite{Chefles,Gong2019}. As each shot is either a click or no-click event, we can use Bernoulli trials to generate analytically this click count distribution after a set number of shots from a suitable no-click event probability.

For classical and quantum illumination there will be a separate click count probability distribution for when an object is present or not. Therefore determining whether or not an object is there depends on the analysis of these probability distributions. The quantum illumination click count probability distribution differs from classical illumination click count probability distribution as there are two data streams for quantum illumination: click counts incident on the signal detector when the idler detector fires and click counts incident on the signal detector when the idler detector does not fire. Also, within a number of trials, there will be a combination of idler-firing and not-firing events, depending on the probability of a click on the idler detector $P(1)_{\textrm{I}}$. The quantum illumination click count probability distribution is a 2D contour plot for any given number of idler firing events.
\begin{figure}
    \centering
    \includegraphics[width=\linewidth]{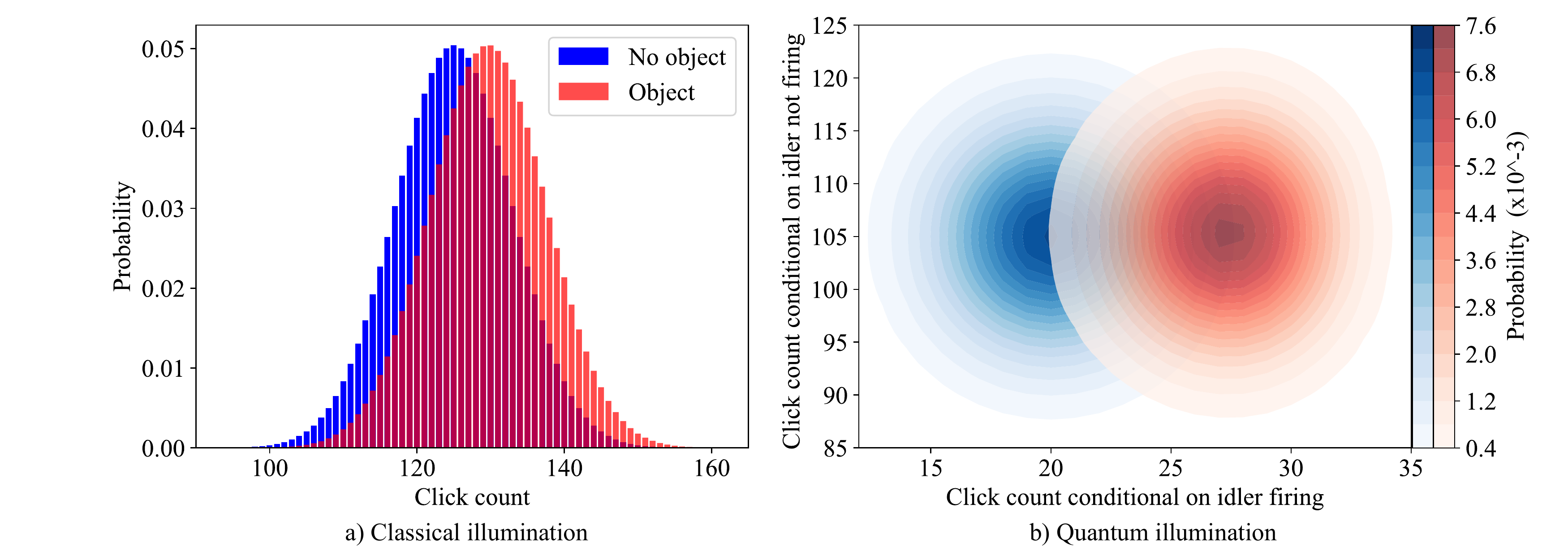}
    \caption{For both sub-figures blue represents the object absent distribution and red represents the object present distribution, with distributions generated from $250$ shots of the system, mean photon number of the signal $\bar{n}_{\textrm{S}}=0.2$, quantum efficiency of all detectors $\eta=0.95$, object reflectivity $\xi=0.707$, thermal background noise mean photon number for the classical illumination detector and signal detector $\bar{n}_{\textrm{B,S}}=1$ and idler detector dark noise mean photon number $\bar{n}_{\textrm{B,I}}=0.01$. (a) Click count distribution for classical illumination-based LIDAR. (b) Click count distribution contour plot for quantum illumination-based LIDAR after $40$ idler firing events and $210$ idler not firing events, where the colour bars represent probability for each distribution.}
    \label{fig:my_label}
\end{figure} Due to the mismatch of the number of click count distribution data streams for classical and quantum illumination, comparing both systems based on the click counts is not simple. Hence, condensing the information from the click counts into one metric will allow for direct comparison between classical and quantum illumination — the log-likelihood test that we now consider performs just that.

\section{LOG-LIKELIHOOD TEST}
\label{llt}

\begin{figure}
    \centering
    \includegraphics[width=\linewidth]{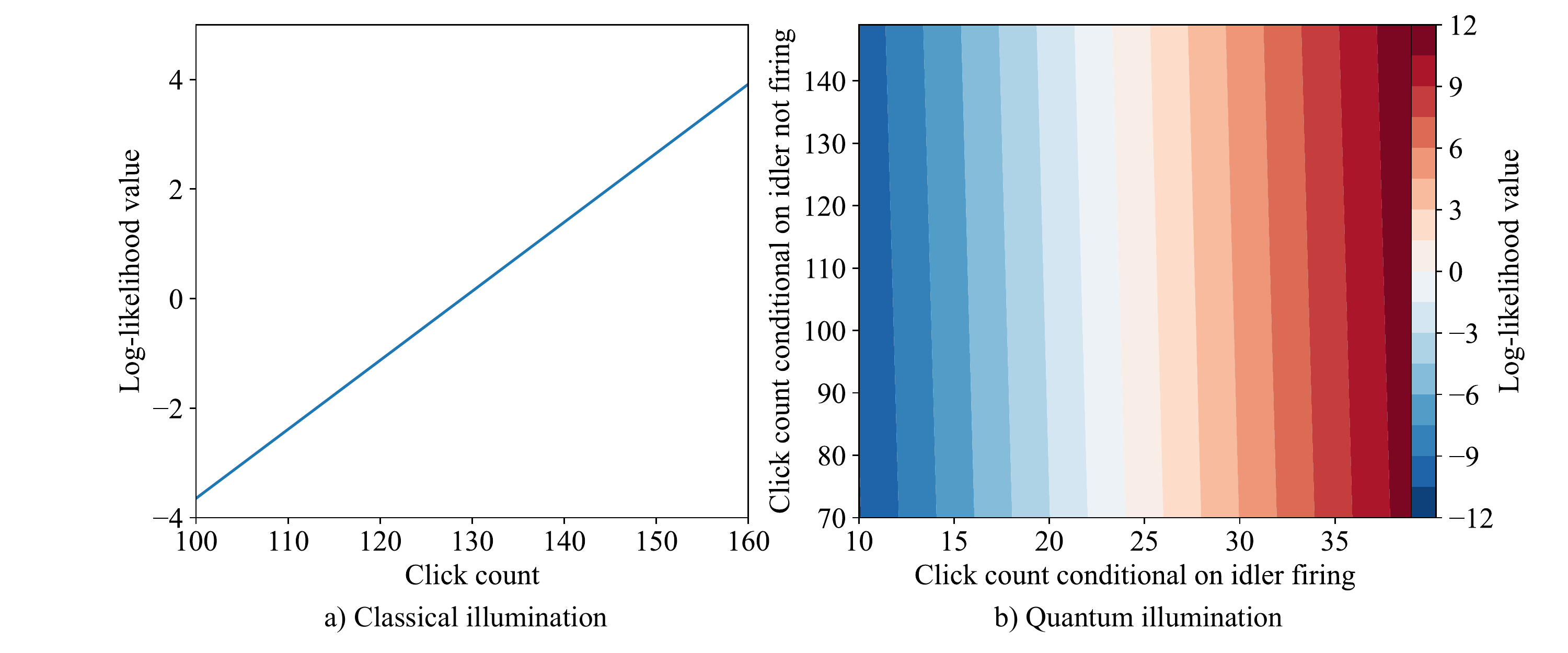}
    \caption{The log-likelihood values displayed for classical illumination (a) and quantum illumination after $40$ idler firing events and $210$ idler not firing events (b). Mean photon number of the signal is $\bar{n}_{\textrm{S}}=0.2$, quantum efficiency of all detectors is $\eta=0.95$, object reflectivity is $\xi=0.707$, thermal background noise mean photon number for the classical illumination detector and signal detector is $\bar{n}_{\textrm{B,S}}=1$ and idler detector dark noise mean photon number is $\bar{n}_{\textrm{B,I}}=0.01$. The number of shots that generated the distributions of click counts is $250$.}
    \label{fig:comblike}
\end{figure}

In order to solve the issue of a mismatch of data streams for click count information between classical and quantum illumination-based LIDAR systems, we use the log-likelihood test, because it transforms click count probability values into a single ratio of the probability of an object being present to the probability of there being no object present. This then allows direct comparison between the two systems. The log-likelihood values for classical illumination and quantum illumination, transform the click count values into
\begin{equation}
    \Lambda(\mathbf{x})=\text{ln}\left(\frac{f_1(\mathbf{x})}{f_0(\mathbf{x})}\right),
\end{equation}
where $\mathbf{x}$ is a vector consisting of the click count information, $f_1$ is the object present click count probability distribution, and $f_0$ is the object absent click count probability distribution. Now, it can clearly be seen that use of the log-likelihood test for quantum illumination condenses the two data streams of click counts into one value. For both quantum and classical illumination, the log-likelihood test implies
\begin{align}
\begin{split}
    \Lambda>0&:\;\text{object presence more likely,} \\ \Lambda=0&:\;\text{neither case more likely, we know nothing,} \\ \Lambda<0&:\;\text{object absence more likely}.
\end{split}
\end{align} 
In Fig. \ref{fig:comblike} we show, the positive log-likelihood values indicate that the presence of an object is far more likely, the log-likelihood value equalling to zero indicates that we know nothing, and the negative log-likelihood values indicate the absence of an object is more likely — all confirmed by visual inspection of the classical and quantum illumination click count distributions.
 It can be seen that quantum illumination has a larger span of log-likelihood values than classical illumination, indicating a greater degree of distinguishability. Also, for quantum illumination while the idler firing events provide the bulk of the distinguishability between possible incident states there is information to be gained from idler not firing events, visualised by the vertical slant in the contour lines. This additional information from the idler not firing events has not been considered in detail in the prior analysis of quantum illumination with simple detection\cite{HYangSPIE}. 

In order to simulate a dynamic system, a rolling-window of click counts for both classical illumination and quantum illumination-based LIDAR is used, where after an initialisation stage, each subsequent iteration of the simulation includes the click count information of the most recent iteration at the expense of the earliest iteration being removed from the click count. This means, for each iteration, there will be click counts from a number of shots equal to the rolling-window size. Following this, the click count information for each iteration is transformed into its corresponding log-likelihood value. We use this to perform the signal analysis of our LIDAR models.
\begin{figure}
    \centering
    \includegraphics[width=12cm]{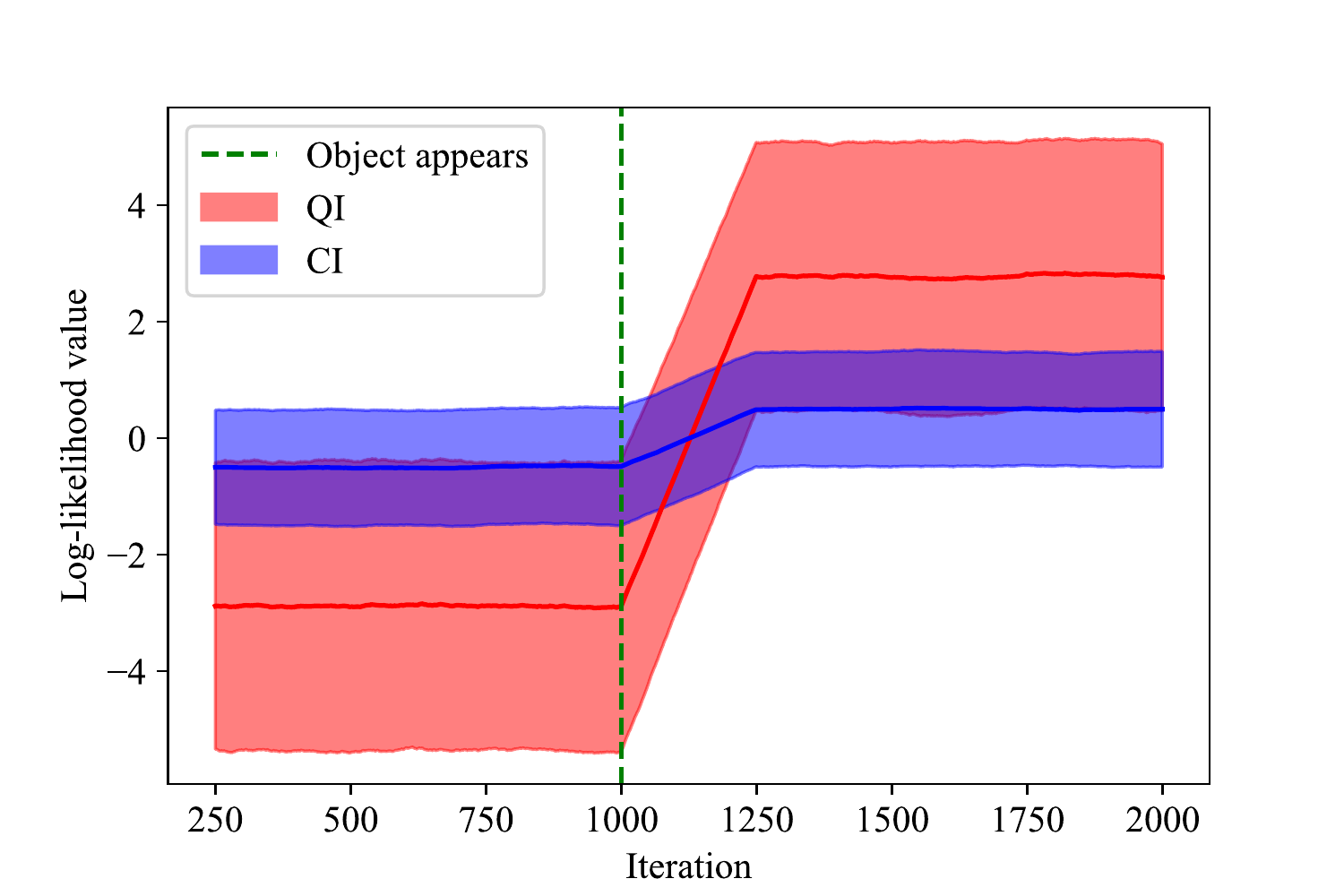}
    \caption{Simulated log-likelihood distributions for classical (CI) and quantum (QI) illumination. Log-likelihood distribution means are the blue and red lines. Shaded regions correspond to one standard deviation around the mean. The rolling-window size is set as $250$. Model parameters: signal mean photon number $\bar{n}_{\textrm{S}}=0.2$, quantum efficiency of all detectors $\eta=0.95$, object reflectivity $\xi=0.707$, thermal background noise mean photon number for the signal detector $\bar{n}_{\textrm{B,S}}=1$ and idler detector dark noise mean photon number $\bar{n}_{\textrm{B,I}}=0.01$.}
    \label{fig:step_log}
\end{figure}
In Fig. \ref{fig:step_log} it is clear that quantum illumination has a far more distinct change in log-likelihood value than classical illumination in a scenario where an object suddenly appears at the $1000^{\textrm{th}}$ iteration of a simulated signal. It can also be seen that it takes the rolling-window size in iterations to fully update from one regime to another. Classical illumination with a single-mode coherent state has not been plotted as it is highly similar to the single-mode thermal state used here. Characterising this change in log-likelihood test values between the cases of object present and not present is what underpins the distinguishability measure $\phi$ that will quantify the performance of quantum and classical illumination.

\section{DISTINGUISHABILITY MEASURE}
\label{dm}
\begin{figure}
    \centering
    \includegraphics[width=17cm]{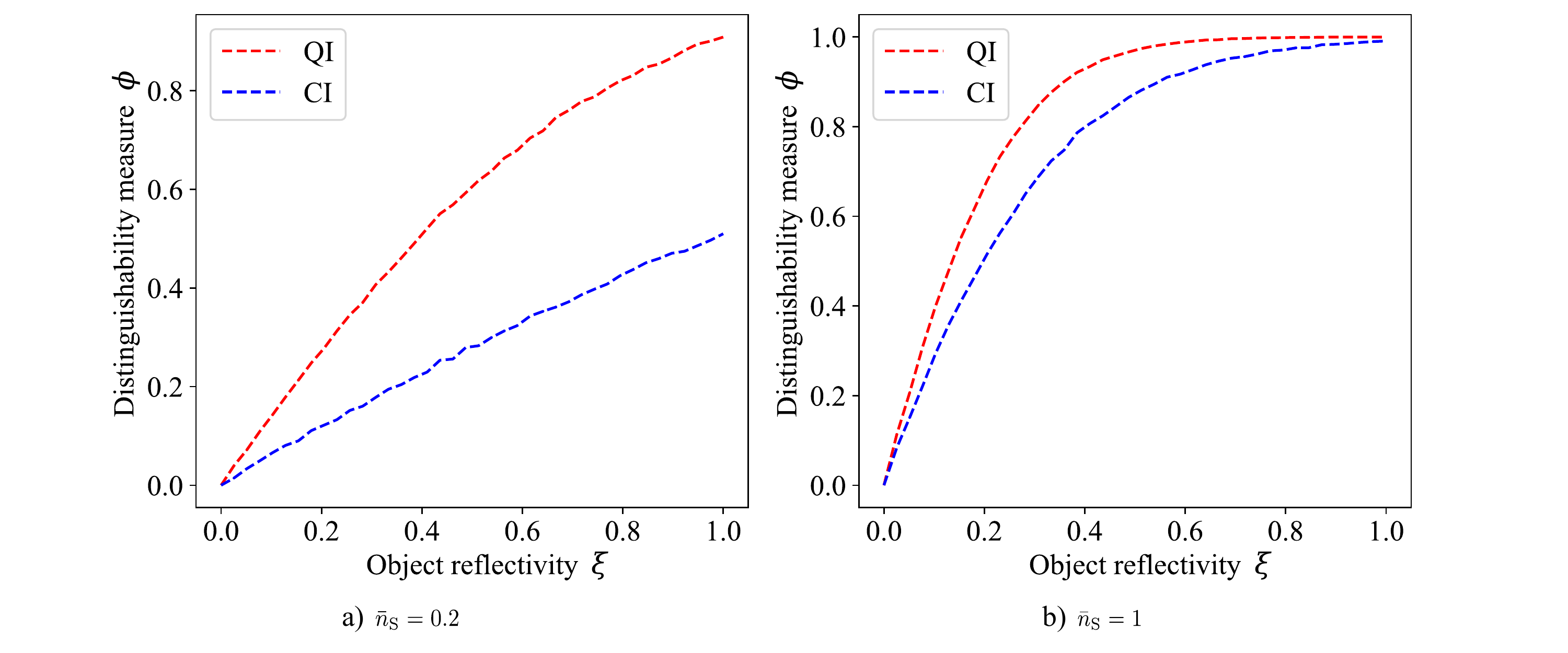}
    \caption{Distinguishability measure $\phi$ plotted against object reflectivity $\xi$, for both quantum (QI) and classical (CI) illumination. The rolling-window size is set as $250$ shots, quantum efficiency of all detectors $\eta=0.95$, object reflectivity $\xi=0.707$, thermal background noise mean photon number for the signal detector $\bar{n}_{\textrm{B,S}}=1$ and idler detector dark noise mean photon number $\bar{n}_{\textrm{B,I}}=0.01$. In a) and b) the mean photon number of the signal $\bar{n}_{\textrm{S}}$ is 0.2 and 1 respectively.}
    \label{fig:performance}
\end{figure}

To characterise the performance of both a classical illumination and a quantum illumination-based LIDAR, a distinguishability measure $\phi$ is constructed from the overlap of the Monte-Carlo simulation generated log-likelihood distributions. The measure $\phi$ used here is defined as \begin{equation}
\phi=1-\int^0_{-\infty}\Lambda_1-\int^\infty_0 \Lambda_0,    
\end{equation} where $\Lambda_1$ is the simulated log-likelihood distribution when an object is present and $\Lambda_0$ is the simulated log-likelihood distribution when an object is absent. The complete overlap of the object present or not distributions corresponds to $\phi=0$ and $\phi=1$ corresponds to these distributions being completely distinct. Usage of the log-likelihood distributions is only reasonable in the limit of a large number of simulation runs, to reduce the inherent noise in the Monte-Carlo simulation process. Figure \ref{fig:performance} shows the distinguishability measure $\phi$ plotted against object reflectivity $\xi$ for both classical illumination and quantum illumination. In both parameter regimes shown there is a quantum advantage, which is more pronounced when there is a weak signal compared to the environmental background noise.

\section{DISCUSSION}
We have presented a model of a LIDAR system applicable to both classical and quantum systems, where we are using simple detectors that can only return a click-event or not, for each shot of the experiment. We have run simulations of this model to generate distributions of click counts after many shots, produced by light incident on the signal detector when the target object is present and not, including previously discarded idler not firing events to make full use of all information available. Following this, we used a log-likelihood to help distinguish the click count distributions, where each combination of click counts at the signal and idler (for quantum) detectors has a corresponding log-likelihood value. The log-likelihood test distinguishes between the presence or absence of the target object. We then run a simulation of an incoming signal when an object is present or not, noting the resultant log-likelihood value. This simulation is repeated until there is a distribution of log-likelihood values for both cases when an object is present or not. A distinguishability measure for the log-likelihood value distributions has been proposed, and it has been used to demonstrate that the log-likelihood value distributions are more distinct for quantum illumination. Hence, quantum illumination has better object detection performance than classical illumination. Our object detection protocol will be developed further to include time-of-flight information, in order to provide range distance of a possible target object. The temporal gating of click-events when tracking an object with time-of-flight information could further improve the performance of quantum illumination.

\acknowledgments 
We would like to thank the UK Ministry of Defence for funding this work and the Defence Science and Technology Laboratory.
\bibliography{main} 
\bibliographystyle{spiebib} 

\end{document}